\documentstyle[12pt,epsfig,color]{article}
\newcommand{\spacing}[1]{\renewcommand{\baselinestretch}{#1}\large\normalsize}
\spacing{1.5}
\begin{document}

\begin{center}
{\large\bf{Exploring sQGP and Small Systems}} \\
\bigskip

{\small
Debasish Das$^{a}$\footnote{email : debasish.das@saha.ac.in,dev.deba@gmail.com}, 
\medskip

$^a$Saha Institute of Nuclear Physics, HBNI, 1/AF, Bidhannagar, Kolkata 700064, India\\
}

\end{center}
\date{\today}
\begin{abstract}

  A strongly interacting Quark-Gluon Plasma (sQGP) is created in the high energy heavy ion
  collisions at RHIC and LHC. Our present understanding of sQGP as a very good liquid with
  astonishingly low viscosity is reviewed. With the arrival of the interesting results
  from LHC in high-energy p+p and p+A, a new endeavour to characterize the transition from these
  small systems to heavy ions (A+A) is now in place, since, even the small systems showed
  prominent similarities to heavy ions in the rising multiplicity domains. An outlook
  of future possibilities for better measurements is also made at the end of this 
  brief review.

\end{abstract}

\section{\bf{Introduction}}

We all know that the normal nuclear matter is made up of protons and neutrons,
which in turn are made up of the quarks~\cite{Riordan:1992hr}
and gluons~\cite{Ellis:2014rma}. The quarks and gluons
are confined inside the colorless particles called hadrons and free colored
particles do not occur. As explained by Quantum Chromo-Dynamics (QCD),
the strong interaction is the governing interaction in the
subatomic world~\cite{Marciano:1977su,Marciano:1979wa}.
One of the important experimental observations that QCD needs to
decipher, is the confinement of the quarks and gluons~\cite{Bruno:2014rxa}. The confinement
property is yet not fully understood, even though qualitatively we know
about the hadron properties (mesons are bound states of a quark and anti-quark
and baryons are bound states of 3 quarks) from the quark model~\cite{Richard:2012xw}.
The refinements of the quark model of hadrons and the development of
QCD, naturally led to expectations that matter
at very high densities~\cite{Berges:1998nf,Kogut:2002kk,Fukushima:2011jc,Wilczek:1999ym}
may exist in a state of quasi-free quarks and gluons,
the Quark-Gluon Plasma (QGP)~\cite{Kisslinger:2014uda,Shuryak:1980tp,Bass:1998vz}.

The very early universe was different than the present times. It was too hot and dense to allow
the quarks and gluons to form hadrons and was apparently filled with a thermalized plasma of
deconfined quarks, anti-quarks, gluons and leptons, which was the primordial QGP~\cite{Ellis:2005xq}.
The universe may have left the QGP phase after a few microseconds with the available quarks and gluons
combining towards the formation of the mesons and baryons~\cite{Heinz:2008tv}.
In our laboratories, we can probe the
QGP~\cite{Adams:2005dq,Adcox:2004mh,Back:2004je,Arsene:2004fa,Abelev:2014ffa,Krintiras:2020imy,Armesto:2015ioy}
which is a deconfined system of
quarks and gluons, by colliding heavy nuclei at relativistic
energies. Such collisions, create QGP which can be
characterized by colored partons as the dynamic degrees of
freedom~\cite{Busza:2018rrf}. Smashing heavy ions, typically Au or Pb ions,
at relativistic energies in the present accelerator facilities, such as
the Relativistic Heavy Ion Collider (RHIC) and
the Large Hadron Collider (LHC), can create the QGP. 
The dynamics of the early universe in terms of the ``Big Bang'' can be studied
experimentally by relativistic nucleus-nucleus collisions at RHIC and LHC
in terms of ``little bangs'' in the laboratory~\cite{Ellis:2005xq,Heinz:2008tv,Busza:2018rrf}.

The main epochs~\cite{Busza:2018rrf} for such a ``little bang'' collision are : (1) the two nuclei
which are lorentz contracted and now disk-like approach each
other and collide with a very small traversal time ($\ll$ 1 fm/c).
(2) The interactions start developing when the two nuclei hit each
other and after such an impact the ``hard''
processes~\cite{Kisslinger:2014uda,Matsui:1986dk,Satz:1983jp,Foka:2016zdb} i.e those which
comprise of relatively large transferred momenta Q $\gg$ 1 GeV between
the quarks, anti-quarks or gluons (partons) inside
the nucleons of the two nuclei produce secondary partons
with large transverse momenta $p_{T}$~\cite{He:2014cla}. During these times
the matter is out of equilibrium and hence will need some
time to equilibrate~\cite{Heinz:2008tv}. (3) The ``soft'' collisions or those with small momentum exchange
Q $<$ 1 GeV cause copius production of particles after sometime
and thermalize the QGP after about 1 fm/c~\cite{Nagle:2018nvi}. The QGP
now expands hydrodynamically and then cools down approximately adiabatically~\cite{Bass:1998vz,Heinz:2008tv}.
(4) The QGP then converts to a gas of hadrons and the hadrons
continue to interact quasi-elastically, further accelerating the expansion
and cooling the fireball until thermal freeze-out ( after $\approx$ 5-10 fm/c) into
thousands of hadrons. The unstable hadrons decay and the stable decay products fly out
to the large scale detectors surrounding the interaction region. During the
hadronization process the chemical composition of the hadron gas is fixed
and remains basically constant afterwards~\cite{Heinz:2008tv,Busza:2018rrf,Nagle:2018nvi}.

By studying the behavior of the matter created in ``little bangs'' we
can explore the phase structure of the strongly
interacting matter~\cite{Ellis:2005xq,Heinz:2008tv,Busza:2018rrf,Nagle:2018nvi}.
The QGP reveals emerging collective behavior~\cite{Nagle:2018nvi,Voloshin:2008dg,Snellings:2011sz,Foka:2016vta}
that originates from the many-body interactions in QCD.
The heavy-ion experiments have explored the close to
perfect fluidity aspects of QGP~\cite{Shuryak:2003xe,Gyulassy:2004zy,Shuryak:2008eq,Lacey:2005qq}, precisely,
with varied experimental observables~\cite{Becattini:2021lfq}. The new state of strongly
interacting matter created in these collisions, have low shear
viscosity($\eta$) to entropy density($s$) ratio, $\eta/s$,
which is close to a nearly perfect
fluid~\cite{Heinz:2008tv,Becattini:2021lfq,Lacey:2006bc,Lacey:2009xx,Bagoly:2015ywa,Jia:2010zza,Lacey:2010ej}.

The paper is organised to start with a brief introduction of QGP
and in Section 2 we have a brief survey of the different avenues of the
formation and promulgation of the strongly coupled QGP.
The term ``strongly coupled QGP (sQGP)'' was
coined~\cite{Shuryak:2003xe,Liao:2006ry} as we have realised
that QGP formed in relativistic heavy ion collisions
is not a weakly coupled gas but on the other hand is more a strongly
coupled liquid~\cite{Gyulassy:2004zy,Liao:2006ry,Nagle:2006cj}.
The realization that QGP created at RHIC is not a weakly coupled gas
but a strongly coupled liquid has aroused a significant development
in this research field. In Section 3 the varied probes for
this dense matter formed in our laboratories and their inferences
towards the understanding of the small systems like p+p and p+A collisions
are discussed. Without the critial understanding of such small systems we
cannot characterize the A+A collisions. Finally, we summarise by looking into
the future scope for such studies that lie ahread.


\section{\bf{sQGP}}
\label{sec:sQGP}

The results from the relativistic heavy ion collision experiments have
changed the theoretical understanding of the properties of the QCD matter.
Also significant know-how has evolved regarding the deconfined QCD matter created
in the central interaction volume at such high energies. Previously QGP
was felt to be a weakly interacting system of quarks and gluons which
might be described by perturbative QCD (pQCD). However contrary to
the expectations, the experimental results
from RHIC~\cite{Adams:2005dq,Adcox:2004mh,Back:2004je,Arsene:2004fa},
have shown that a hot, strongly interacting, nearly perfect
and almost opaque relativistic liquid, also
termed as the strongly coupled QGP was created in central Au+Au collisions
at the top RHIC energy regime~\cite{Heinz:2008tv,Shuryak:2003xe,Gyulassy:2004zy,Shuryak:2008eq}.

The  comparative studies of the experimental data~\cite{Adams:2005dq,Adcox:2004mh,Back:2004je,Arsene:2004fa},
and especially the elliptic flow ($v_{2}$)~\cite{Lacey:2005qq}, in terms of the
hydrodynamic models showed the nearly perfect fluid behavior
of QGP. Such inferences indicate that its properties correspond
to non-perturbative, strongly interacting matter.
RHIC results showed that the resulting plasma could be well
described by a hydrodynamic picture of a nearly ideal liquid,
which show very limited internal friction or in other words very small shear
viscosity ($\eta$). The created medium in such relativistic collisions, can connect
to the pressure gradients by flowing apparently unobstructed~\cite{Becattini:2021lfq,Shuryak:2005pp,Thoma:2004sp}.

Shear viscosity, $\eta$, is a characterizing parameter
for fluids~\cite{Shuryak:2005pp,Hirano:2005wx,Schafer:2009dj,Schafer:2009kj}
and can be defined in terms of the friction force $F$ per unit
area $A$ produced by a shear flow with transverse flow gradient $\nabla_{\! y} v_x$,
\begin{equation}
\label{eta_one_defn}
\frac{F}{A}=\eta\, \nabla_{\! y} v_x\, .
\end{equation}
Small shear viscosity is a benchmark for a good fluid.

Shear viscosity for a weakly coupled gas can be estimated as 
\begin{equation}
\label{eta_two_eta_exp}
\eta = \frac{1}{3}\,n p \lambda\, ,
\end{equation}
where $n$ is the density, $p$ is the average momentum of the
gas molecules, and $\lambda$ is the mean free path. The mean free
path can be expressed as $\lambda = 1/(n\sigma)$ where $\sigma$
is a preferable transport cross-section. For relativistic fluids
it is more natural to normalize $\eta$ to the entropy density $s$  
rather than the particle density $n$. 

It has been observed that good fluids are characterized
by $\eta/s$ $\sim$ $\hbar/k_{B}$ and this  value is consistent
with simple theoretical propositions. For all fluids, the proposed
lower bound based on the results from string theory~\cite{Kovtun:2004de}, is,

\begin{equation}
\label{eq_three_eta_s_limit}
\frac{\eta}{s} \geq \frac{\hbar}{4 \pi k_{B}}.
\end{equation}

A ``perfect fluid'' saturates around this value by dissipating
the smallest possible amount of energy. A perfect fluid thus
follows the laws of fluid dynamics in the largest
possible domain~\cite{Schafer:2009dj,Schafer:2009kj,Kovtun:2004de}.

The experimental results from RHIC indicate that the matter produced in
nuclear reactions has a small ratio of $\eta/s$~\cite{Lacey:2006bc,Lacey:2009xx}.
The discovery of such a close perfect fluid nature established
relativistic fluid dynamics as the new frame-work for deciphering the
bulk evolution of the system~\cite{Becattini:2021lfq,Shuryak:2005pp}.
The observations illustrate that QGP near $T_{c}$ is a strongly coupled one with
the properties of a liquid with very low
viscosity rather than that of a dilute gas~\cite{Niemi:2011ix,Asakawa:2006tc}.

Analysis infers~\cite{Lacey:2010ej} that the averaged
specific viscosity of the QGP produced in LHC collisions
is quite similar to that for the dense matter created in RHIC energy
domain. So, the domain in which matter produced at RHIC/LHC is,
$T_{c}$ $<$ $T$ $<$ $2T_{c}$,  was  renamed  into
a strongly coupled QGP or ``sQGP'' in
short~\cite{Lacey:2005qq,Berrehrah:2014ysa,Csanad:2012mp,Rapp:2008fv,Wang:2009bn,Shuryak:2014zxa,Heinz:2005zg}.
On the other hand the low value for $\eta/s$ could also result from an
anomalous viscosity $\eta_{A}$, originating from turbulent
color magnetic and electric fields dynamically produced
in the expanding quark-gluon plasma~\cite{Lacey:2006bc,Asakawa:2006tc}. That is,

\begin{equation}
\label{eq_four_eta_add}
1/\eta = 1/\eta_{A} + 1/\eta_{C} ,
\end{equation}

where $\eta_{A}$ subjugates over the collisional viscosity $\eta_{C}$.
Such arguments do not rule out a more complex structure of the
gluonic component of the matter produced in
the relativistic collisions~\cite{Liao:2016nqa}.

At LHC energies the inital energy density(at $\tau_{0}$ = 1 fm/c)
is about 15 GeV $fm^{-3}$~\cite{Muller:2012zq}. It is approximately
a factor of three higher than the Au+Au collisions at the highest
energy regime at RHIC. Some researchers expected that the QGP
produced at the LHC would turn back to the previous picture, where quarks and
gluons were more weakly coupled at higher temperature. Then the mean free path
of particles in the medium and the viscosity will be significant.
As a result the experimental signature will emerge as
smaller flow components($v_{n}$). But the ALICE elliptic
flow $v_{2}$ results~\cite{Aamodt:2010pa} have clearly shown,
the opposite. The dependence of $v_{2}$ on transverse momentum
is comparable with the RHIC measurements and ALICE
has also established that radial flow grows with energy.

Understanding sQGP was a challenge which we have researched
from RHIC data. However the LHC program has added a lot to our understanding,
and the paramount issues in the field now include a critical search
to study the evolution between p+p, p+A collisions which are known as ``small systems''
and heavy ion A+A collisions, with an goal to understand ``the smallest drops''
of the sQGP showing collective/hydrodynamics behavior~\cite{Shuryak:2014zxa}.
Some of these assumptions are getting tested
and understood carefully both in RHIC~\cite{PHENIX:2018lia} and
LHC~\cite{Khachatryan:2015waa,Khachatryan:2016txc} experiments.

At LHC since the collision energies increase,
one expects a QGP which is hotter. Such favourable high energy
of LHC is more evident in the area of parton energy loss
analogous to the opaque nature of the sQGP where the kinematic
domain exceeds that of RHIC. The significant impact
of this increase of the collision energy is the huge excess
of the rates of hard probes, such as jets, electro-weak particles
and heavy-flavors, including the full family of
quarkonia ( $c\bar{c}$ and $b\bar{b}$ bound states)~\cite{Muller:2012zq,Cacciari:2011mb}.
With a larger in-elastic cross-section, the production of $b\bar{b}$ pairs
will increase more in LHC energies. The abundance of $b\bar{b}$ pairs
enable the possibility for bottom quark and anti-bottom quark pairs to recombine,
following bottomonium state breakup, or combination after the pair forms
from the open bottom states. The available high rates allow detail studies of
the dense medium using the interactions of these probes with the
medium constituents~\cite{Cacciari:2011mb,Das:2018xel,Andronic:2015wma}.

The elastic re-scattering of the heavy quarks in the sQGP is an
important element for the understanding of heavy-flavor and single-electron/muon
observables in heavy ion reactions at collider energies~\cite{He:2014cla,Das:2011bj}.
The produced heavy-flavor interacts with the dense medium
by exchanging energy and momentum. The ratio of the measured number of
heavy-flavors in heavy ion (A+A) collisions to the expected number in the
absence of nuclear or partonic matter i.e p+p collisions,
is the definition of nuclear modification factor($R_{AA}$) which
is suppressed at high transverse momentum~\cite{Andronic:2015wma}.
The elementary degrees of freedom and basic forces at the shortest
distances are understood via small systems~\cite{Heinz:2008tv}.
So a clear understanding of the small systems emerge as a necessity. 
The small collision systems like p+p and p+A collisions at LHC energies
thus needs detail study to understand the initial and
final state effects in Cold Nuclear Matter(CNM), which can provide
baseline for the interpretation of
heavy ion (A+A) results~\cite{Nagle:2018nvi,Das:2021nqw}.


\section{\bf{Small Systems}}
\label{sec:small_systems}

Study of QGP requires reference measurements which is provided by
the small system (p+p and p+A) collisions~\cite{Nagle:2018nvi,Das:2021nqw}.
QGP is not expected to be formed in small systems as the transverse size
of the overlap region is comparable to that of a single
proton~\cite{Abelev:2014ffa,Abelev:2012qh,Ni:2018spv,MoreiraDeGodoy:2017wks}.
Particle production in A+A and p+A, as compared to p+p collisions,
expressed as $R_{AA}$, is termed as the nuclear modification factor.
It has long been formulated to understand particle production mechanisms~\cite{Andronic:2015wma}.
The $R_{AA}$ of heavy-flavor is expected to be less suppressed and
elliptic flow $v_{2}$ of heavy-flavor is felt to be smaller in comparison
with the light hadrons. The experimental results from ALICE, however, show that
the suppression of heavy-flavor hadrons (D-meson) at high transverse
momentum ($p_{T}$) and its elliptic flow $v_{2}$ are comparable to those of
the light hadrons~\cite{ALICE:2012ab,Abelev:2013lca}, which needs to be
understood~\cite{Dong:2019byy}. Hence looking into the p+A collisions is required~\cite{Acharya:2018yud},
where medium absence provides necessary conditions, to isolate the nuclear effects
from the initial hard-scattering processes which we often describe
as CNM~\cite{Vogt:2010aa,Fujii:2013gxa,Vogt:2015zoa,Albacete:2017qng}.

Broadly the CNM effects emcompass : (i) initial-state nuclear
effects on the parton densities (i.e shadowing); (ii) coherent energy  loss
comprising of initial-state  parton energy loss  and  final-state energy loss;
and (iii) the final-state absorption by nucleons, which is expected
to be negligible at LHC energies. The CNM effects like the change of the Parton Distribution
Functions (PDFs) within the nucleons contained within the nuclei, as compared
to the unbound nucleons can modify the interaction and production cross-sections~\cite{Armesto:2018ljh}.
That's why the p+A collisions are important to decouple the effects of QGP from those of CNM,
and to provide very much required input to the understanding of
A+A collisions~\cite{Vogt:2010aa,Fujii:2013gxa,Vogt:2015zoa,Albacete:2017qng}.
The nuclear modification factor of charged particles from CMS experiment~\cite{Khachatryan:2016odn}
in p+Pb collisions, in contrast to the Pb+Pb system at top LHC energies of $\sqrt{s}_{\rm NN}$=5.02 TeV,
demonstrate no suppression in the 2-10 GeV/c $p_{T}$ region.
However we visualize a weak momentum dependence for $p_{T}$ $>$ 10 GeV/c in
the p+Pb system, since we observe a moderate excess above unity
at high $p_{T}$ for charged particles. Also for heavy-flavor(D-meson),
the nuclear modification factor, measured by
ALICE experiment~\cite{Abelev:2014hha} in p+Pb collisions
at same energies, show no suppression within the
uncertainties in the measured $p_{T}$ range of 1-24 GeV/c.
The strong  suppression  of  the  D-meson  yields
for $p_{T}$ $>$ 3 GeV/c has been observed in central and semi-peripheral Pb+Pb collisions~\cite{Adam:2015sza},
whereas, for the charged particles~\cite{Khachatryan:2016odn} we see for
$p_{T}$ $<$ 2 GeV/c a rising trend in both p+Pb and Pb+Pb systems.
In the Pb+Pb collisions the charged particles~\cite{Khachatryan:2016odn}
then show a significant suppression in the 2$<$$p_{T}$$<$10 GeV/c region,
and again a rising trend around 10 GeV/c to the highest $p_{T}$.
The p+Pb and Pb+Pb nuclear modification
factors presented in these papers~\cite{Khachatryan:2016odn,Abelev:2014hha,Adam:2015sza},
covering the light and heavy quarks respectively, provide stringent
constraints on cold and hot nuclear matter effects.
They also clearly establish why the CNM effects are of crucial importance for
accurate interpretation of the measurements in heavy ion collisions and
in turn advocate the necessity of studying the small system collisions.

But at LHC energies do we see any new features in p+p collisions?
At LHC energies the particle multiplicity is high
and even reach values, which are of the same order as
those found in heavy ion collisions at lower energies, and as a matter of fact,
they are well above the ones observed at RHIC for peripheral
Cu+Cu collisions at $\sqrt{s_{\rm NN}}$ = 200 GeV~\cite{Alver:2010ck}.
When LHC started with the p+p collisions, the high-multiplicity
environment revealed a ``ridge'' which was measured by CMS~\cite{Khachatryan:2010gv}
while studying the long-range azimuthal correlations for 2.0 $<$ $|\Delta\eta|$ $<$ 4.8.
The first observation of a long-range ridge-like structure
at the near-side ($\Delta\phi$ $\approx$ 0) was observed for 7 TeV p+p collisions. 
For the high multiplicity domain of $N\approx$ 90 or higher,
this notable feature is clearly observed for large rapidity differences $|\Delta\eta|$ $>$ 2.
Also in the high-multiplicity p+Pb collisions at $\sqrt{s_{\rm NN}}$=5.02 TeV,
the azimuthal correlations for 2.0 $<$ $|\Delta\eta|$ $<$ 4.0 showed a qualitatively similar
long-range structure at the nearside $\Delta\phi$ $\approx$ 0.
Thus the long-range,  near-side angular correlations in
particle production emerged in p+p and subsequently in p+Pb collisions~\cite{CMS:2012qk},
which was further followed by an away-side structure, located at $\Delta\phi$ $\approx$ $\pi$ 
and exceeding the away-side jet contribution, in p+Pb collisions~\cite{Abelev:2012ola,Aad:2012gla}.

In a typical p+p collision, a ridge correlation is not expected
because the system is too dilute to produce a fluid-like state.
This paved the way to encourage the researchers to look for a
detailed investigation of the existence of collective phenomena
in p+p collisions which was known since long in heavy ion collisions~\cite{Abelev:2009af}.
The strong evidence for the collective nature of the long-range correlations was observed
with the charged particles (light quarks)~\cite{Khachatryan:2016txc} by CMS experiment
at $\sqrt{s}$ = 13 TeV. Also the  elliptic flow ($v_{2}$) coefficients
for heavy-flavor decay muons was measured by ATLAS in p+p collisions at same energy~\cite{Aad:2019aol}.

Since heavy quark yields in heavy ion collisions are expected to be modified
relative to minimum bias p+p collisions~\cite{Andronic:2015wma}, the obvious
question arises if their production  rates in high-multiplicity  p+p collisions
at LHC energies show any effect like J/$\Psi$ suppression~\cite{Matsui:1986dk,Satz:1983jp}.
A stronger than linear rise of the relative production
of J/$\Psi$ as a function of multiplicity was observed for
$p_{T}$-integrated yields and this increase is stronger
for high-pT J/$\Psi$ mesons which we see for p+p collisions
at $\sqrt{s}$ = 13 TeV~\cite{Acharya:2020pit}.
An esclation of the relative J/$\Psi$ and $\Upsilon$
yields~\cite{Adamova:2017uhu,Chatrchyan:2013nza,Aaboud:2017cif}
with the relative charged-particle multiplicity was observed in p+Pb
collisions at $\sqrt{s_{\rm NN}}$=8.16 TeV~\cite{Acharya:2020giw}.
The results in p+A are very similar to the results from
p+p collisions~\cite{Chatrchyan:2013nza,Aaboud:2017cif,Abelev:2012rz}.
The rise of the J/$\Psi$ normalized yields are comparable to
the increase observed for D-mesons~\cite{Adam:2015ota,Adam:2016mkz} which
indicate that a common mechanism may be at its origin.
A plethora of new, unexpected phenomena have been observed so far in
small system (p+p and p+A) collisions, which,
produce remarkable similarities to heavy ion phenomenology.

\section{Summary and Outlook}

The more-central Au+Au collisions at RHIC, on the basis of elliptic-flow systematics,
have been characterized in terms of a sQGP with small viscosity — a ``perfect liquid''.
Crucial input to our comprehension of the sQGP were inferred from the measurements
of ``collective flow'', which in other words is the correlated emission of
particles in azimuthal angle around the axis of the colliding beams.
Conventionally, we have diagnosed the effects of the sQGP on the final-state particle production
and correlations in A+A collisions, by using the relative to
baseline measurements of p+p and p+A collisions, and thus assuming
that in the smaller, and therefore shorter-lived systems, no QGP effects can happen. 
At LHC we found new things and even the small systems showed flow features in
the rising multiplicity domains. With the increasing multiplicity,
the p+p and p+A collisions enter the stage  where the
macroscopic description (thermodynamics and hydrodynamics) becomes applicable.
While hydrodynamic models, when applied to p+A data, can explain many of the observed
features, there are serious questions regarding their applicability~\cite{Shuryak:2013ke}.
Thus, a very detailed description of a broad range of signatures, in an even broader range of
systems, will be required to finally demonstrate a full understanding of these new discoveries.

Data which will be collected in Run-3 at the LHC, will be
a significant addition for such studies. Better picture will
be also available with the results from p+Pb collisions.
Also exceptionally high-multiplicity p+p collisions
are expected in Run-3 and 4 at LHC~\cite{Noferini:2018are}.
The LHC delivered nearly 30 $fb^{-1}$ by the end of 2012 and
propose to reach 300 $fb^{-1}$ in its first 13-15 years of operation.
The second long shutdown (LS-2) before Run-3 will consolidate
the luminosity and reliability as well as the upgrading of the LHC injectors.
After LS-3, the machine will be in the High Luminosity configuration.
The High Luminosity LHC(HL-LHC) is an important
and extremely challenging, upgrade~\cite{Apollinari:2017cqg}.
The  large p+p collision  data  sets  expected  to  be  collected  at
the  HL-LHC  will provide a compelling setting
for these investigations~\cite{Noferini:2018are,Chapon:2020heu}.
Such higher multiplicities will help us to bridge the gap between
the p+p and heavy ion collisions, with better detector
upgrades in LHC experiments~\cite{Citron:2018lsq}.


\bibliography{apssamp}

\end{document}